\documentclass[floatfix,preprintnumbers,amssymb,twocolumn,,pre]{revtex4-1}
\usepackage{natbib}
\usepackage{graphics}
\usepackage{booktabs}
\usepackage{amsfonts, amsmath, amsthm,mathrsfs}
\usepackage{mathtools}
\usepackage{wrapfig, tikz,url,float}
\usepackage{dcolumn}
\usepackage{bm}
\usepackage{mathtools}
\usepackage{color}
\def \kc {k_{\rm c}}
\def \Fdtog {\bm{F}^{{\rm d}\to {\rm g}}}

\def \fnot {f_{\rm 0}}
\def \dtwo {d_{\rm 2}}
\def \phim {\phi_{\rm m}}

\def \Np {N_{\rm p}}

\def \XX {{\bm X}}
\def \VV {{\bm V}}

\def \xx {{\bm x}}

\def \del2z {\partial^{2}_{z}}

\def \rhog {\rho_{\rm g}}
\def \rhod {\rho_{\rm d}}

\def \uu {{\bm u}}

\newcommand{\Eq}[1]{Eq.~(\ref{#1})}

\newcommand{\ddt}[1]{\frac{d#1}{dt}}

{}

\def \St  {\mbox{St}}

\def \taup {\tau_{\rm p}}
\def \tauf {\tau_{\rm f}}

\def \kf  {k_{\rm f}}

\def \kf  {k_{\rm f}}

\newcommand{\fig}[1]{Fig.~(\ref{#1})}
\newcommand{\subfig}[2]{Fig.~(\ref{#1}#2)}

\def\drawing #1 #2 #3 {
\begin{center}
\setlength{\unitlength}{1mm}
\begin{picture}(#1,#2)(0,0)
\put(0,0){\framebox(#1,#2){#3}}
\end{picture}
\end{center} }
\def \rhog {\rho_{\rm g}}

\newcommand{\beq}{\begin{equation}}
\newcommand{\eeq}{\end{equation}}

\newcommand{\REM}[1]{{}}

\usepackage[linktocpage,colorlinks=true,linkcolor=blue,citecolor=blue]{hyperref}

\begin{document}
\title{Clustering and energy spectra in two-dimensional dusty gas turbulence}
\author{Vikash Pandey}
\affiliation{TIFR Centre for Interdisciplinary Sciences, Hyderabad.}
\author{Dhrubaditya Mitra}
\affiliation{NORDITA, Royal Institute of Technology and Stockholm University,  Stockholm}
\author{Prasad Perlekar}
\affiliation{TIFR Centre for Interdisciplinary Sciences, Hyderabad.}
\date{\today}
\begin{abstract}

We present Direct Numerical Simulation (DNS) of heavy inertial particles (dust) 
immersed in two-dimensional turbulent flow (gas). The dust are modeled as
mono-dispersed heavy particles capable of modifying the flow through two-way coupling.
By varying the Stokes number ($\St$) and the mass-loading parameter ($\phim$), we study
the clustering phenomenon and the gas phase kinetic energy spectra. We find that the
dust-dust correlation dimension ($\dtwo$) also depends on $\phim$. In particular, 
clustering decreases as mass-loading ($\phim$) is increased. In the kinetic 
energy spectra of gas we show: (i) emergence of a new scaling regime, (ii) the scaling exponent
in this regime is not unique but rather a function of both $\St$ and $\phim$. Using a scale-by-scale 
enstrophy budget analysis we show in the new scaling regime,  viscous dissipation due to the gas balances  
back-reaction from the dust.
\end{abstract}

\maketitle
\section{\label{sec:intro} Introduction}

In nature, turbulent flows often include small particles embedded
within the flow, typical examples are 
(a)  proto-planetary disks (gas and dust)~\cite{Arm10}, 
(b) clouds (air and water droplets)~\cite{Pruppacher2010microphysics} 
and (c) aeolian processes (wind and sand)~\cite{kok2012physics}.
Analytical, numerical and experimental studies of such multiphase
flows have flourished in the last decade~(see e.g. Refs.~\cite{Zai+ali+sin08,tos+bod09,pum+wil16,gus+meh16} for a review). 
For notational convenience, in the rest of this paper, we shall call
the solvent phase ``gas'' and the solute phase ``dust''.
Often the simplest model used to study such multiphase flows
assume that the dust is a collection of 
heavy, inertial particles (HIPs) which do not alter the gas flow.
The equation of motion of the dust particles are

\begin{subequations}
\begin{align}
\ddt{\XX(t)} &= \VV(t), \label{eq:dxdt}\\
\ddt{\VV(t)} &= \frac{1}{\taup}\left[\uu(\XX,t) -\VV  \right] , \label{eq:dvdt} 
\end{align}
\label{eq:HIP}
\end{subequations}
where $\XX$ is the position, $\VV$ is the velocity of a dust particle,  and $\uu$ 
is the velocity of the gas at a point $\XX$. 
For incompressible flows, in addition to the Reynolds number, an
additional dimensionless number appears, the Stokes number
$\St \equiv \taup/\tauf$ where $\tauf$ is a characteristic timescale
of the flow of gas. 
If the size of dust grains are comparable to or larger than the
dissipative scales of the flow then the simple approximation encoded
in \Eq{eq:HIP} is not valid any more. Furthermore, \Eq{eq:HIP} is a reasonable
model of reality if the number density of the dust
grains is so small that both the dust-dust interaction and the
back-reaction from the dust phase to the gas phase can be ignored. 
In this paper, we study the consequences of relaxing this last
assumption. 
One of our motivations is the recent realization that the  dust in
astrophysical plasma cannot be treated merely as a passive component.
In particular, the inclusion of the back-reaction allows for novel 
instabilities, e.g.  the streaming instability~\citep{you+god05,joh+you07}, to 
manifest itself. 

In the absence of dust, the turbulence in the gas phase has been
extensively studied \cite{Fri96,Pop00,Dav04}.
The pioneering work of Kolmogorov~\cite{kol41} has established that
in three-dimensions the (angle-integrated) energy spectrum of the gas shows power-law
behavior $E(k) \sim k^{-5/3}$ within the inertial range followed by
the dissipation range where the energy spectrum shows exponential
decay~\footnote{Experiments and recent
numerical simulations have demonstrated that the Kolmogorov picture
is not complete, but must include corrections due to
intermittency. The intermittency corrections to the energy spectrum
is small and is ignored in this paper.}.
More importantly, the inertial range spectral exponent is universal, i.e. it does 
not depend on the Reynolds number and  the mechanism of turbulence generation.  
Does the presence of dust modifies this energy spectrum?  Obviously, in general, the answer depends on the number, size,
and shape of the dust grains. In this paper, we study this question using direct numerical
simulations (DNS) of the dusty gas flow. 

A recent paper~\cite{gua+bat+cas17} has suggested
that in the presence of dust a new power-law behavior can emerge 
where $E(k) \sim k^{-4}$  in three dimensions.   
Is this exponent universal, in the sense that, is it independent of the
Stokes number and the dust concentration? 
It is difficult to provide an  answer to this question because
an accurate determination of the exponent requires 
obtaining clean scaling of the energy spectrum over at least a
decade. 
This is a formidable task in three dimensions but is a much simpler
proposition in two-dimensions. 
Hence to understand the universality (or lack thereof) we study
this problem in two dimensions. 
 
In two-dimensional gas turbulence \cite{bof+eck12}, forced at large scales (small $k$) and 
in presence of air-drag friction ($\alpha$), the energy spectrum is universal with respect 
to the Reynolds number but does depend on the air-drag-friction 
coefficient \cite{nam00,boffetta_2005}.
The scaling exponent and its non-universality can be understood 
as an effect of the loss of enstrophy due air-drag-friction~\cite{verma2012variable}. 
In our simulations we choose an $\alpha$ such that in the absence of
dust $E(k) \sim k^{-3.9}$. 
We then perform extensive simulations of the dusty-gas flow by varying
both the Stokes number and the mass-loading-parameter
($\phim$, ratio of the total mass of the dust to the total fluid mass). 

The rest of the paper is organized in the following manner. In section 
(\ref{model}) we present our model and describe how it is implemented numerically. 
The result section (\ref{result}) is divided in three subsections. First subsection (\ref{clustering}) is devoted to discussion of the pair distribution function of dust particles where we show that
increasing mass-loading parameter reduces the clustering of dust. In the remaining subsections (\ref{spectra}) and (\ref{budget}), we present energy spectra and 
scale-by-scale enstrophy budget respectively 
for the gas phase. We show that indeed in the presence dust-gas coupling,
a new scaling range emerges in the gas kinetic energy spectra. Furthermore, using a scale-by-scale enstrophy budget analysis  we show that the new scaling appears due to  a balance  between the  injection (from the
dust to the gas) and viscous dissipation.  Our main result is that
the scaling exponent is {\it not universal} but 
depends on both $\St$ and the mass-loading parameter $\phim$. 
Finally, in section (\ref{conclusion}) we conclude the paper. 
 
\begin{figure}
	\begin{center}
		\includegraphics[width=0.8\linewidth]{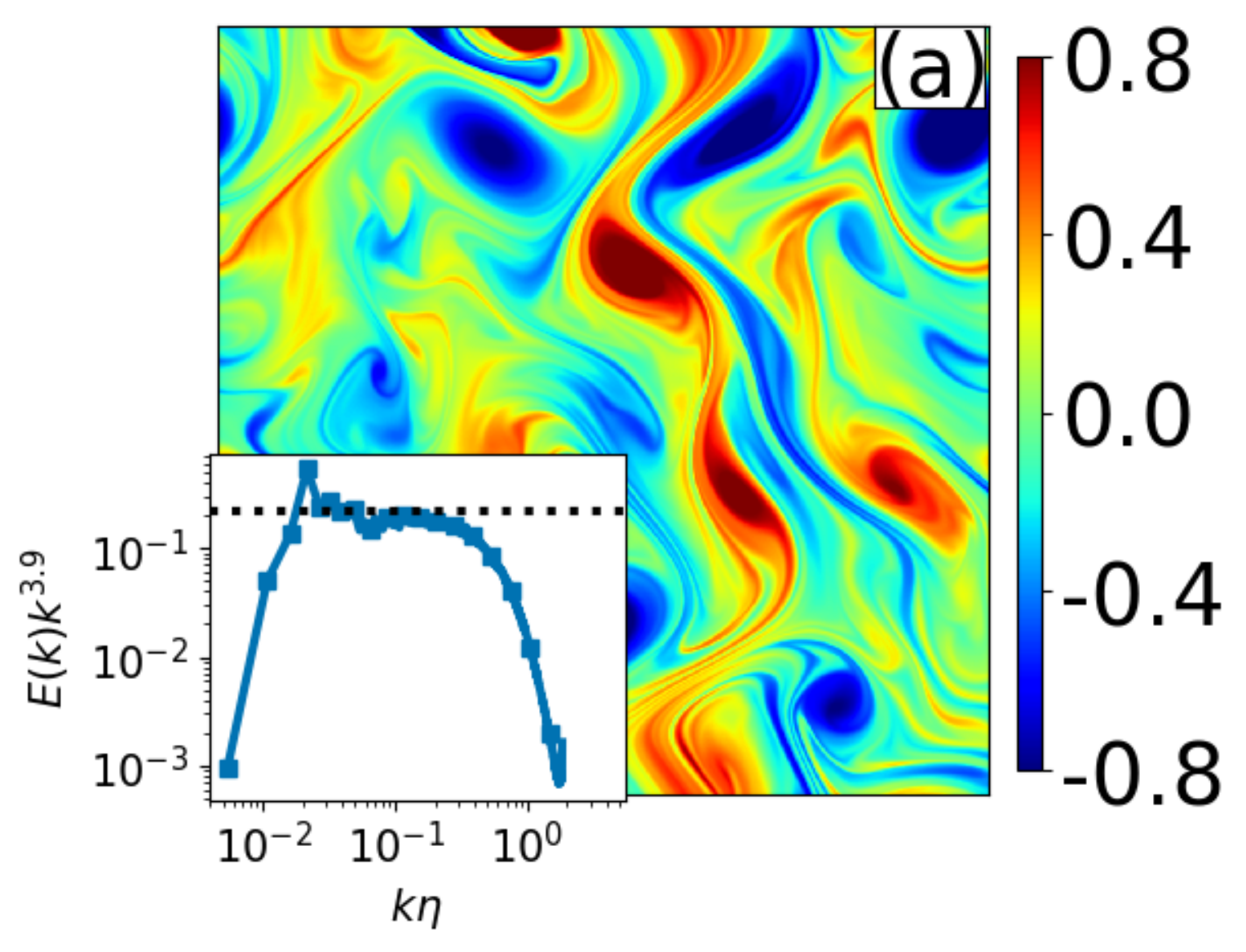}
		\includegraphics[width=0.8\linewidth]{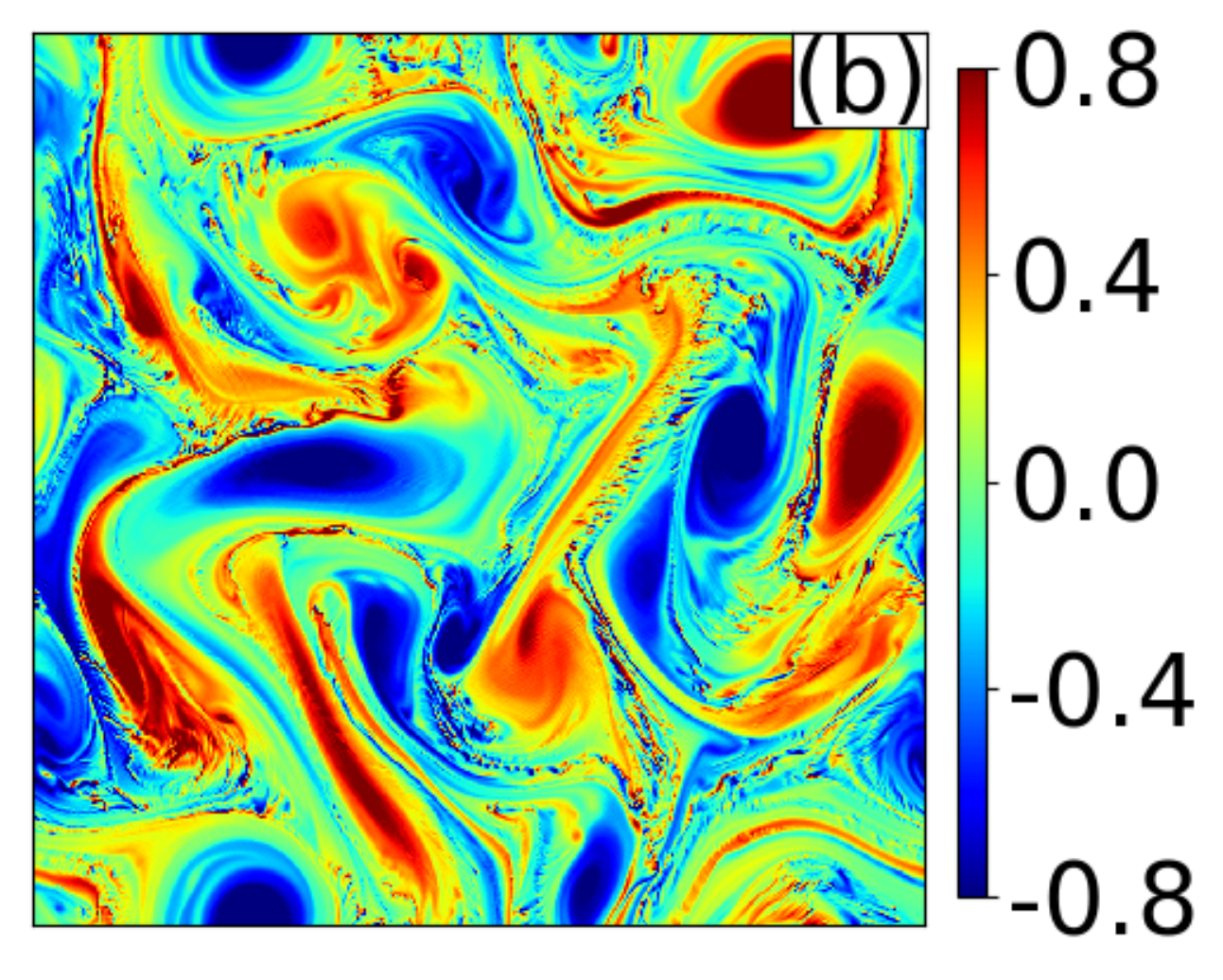}
		\includegraphics[width=0.8\linewidth]{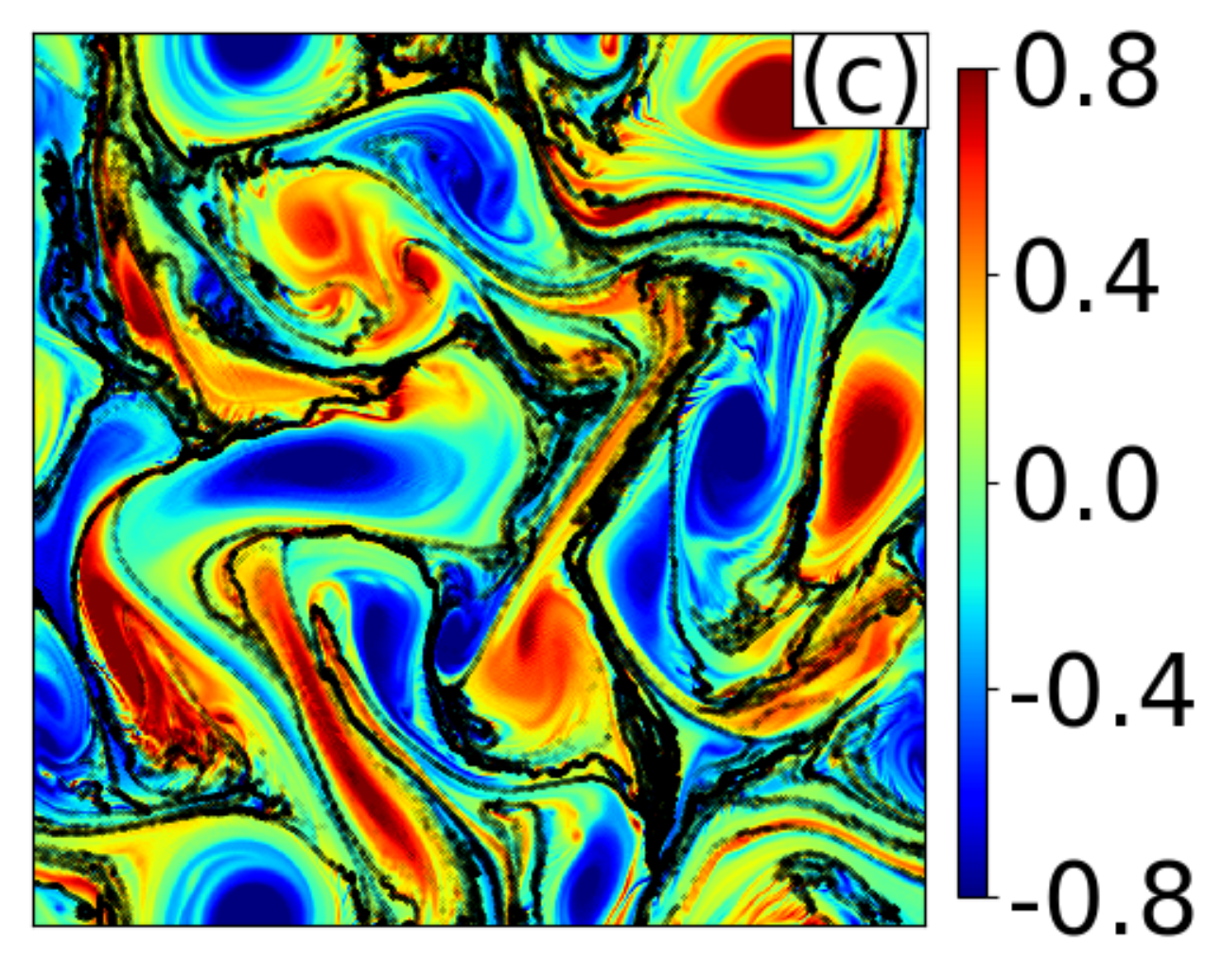}
		\caption{\label{fig:vort}(a) Representative steady-state snapshot of turbulent vorticity
        field $\omega$ from our simulation. (Inset) Log-log plot of compensated energy spectrum
        [$k^{3.9}E(k)$] versus $k\eta$, where $\eta$ is the Kolmogorov dissipation length scale.
        To generate turbulent flow we take $\alpha = 10^{-2}$, $f_0 = 5\times10^{-3}$, $\kf = 4$
        and $\nu = 10^{-5}$ units. For the resulting turbulent flow,
        $\eta = 5.4\times 10^{-3}$ units,
		 $\tau_\eta = 2.89$ units, and enstrophy dissipation rate $\beta =
         2.8\times10^{-4}$ units.
		(b) Representative snapshot of $\omega$ during steady state for $\St=0.33~{\rm and}~
        \phim=1.0$. (c) The positions of all the dust particles are overlaid as black dots in
        underlying vorticity plot of (b). The diameter of each dust particle is $\sim 0.1\eta$.
        We take $N=1024$ for all the simulations in section (\ref{clustering}) and $N=4096$ for
        simulations in sections
		(\ref{spectra}) and (\ref{budget}). We vary $\St$ in the range $0.17-1.67$ and $\Np$ in
        range $1.5\cdot10^{4}-1.5\cdot 10^{5}$ to achieve mass-loading ($\phim$) of $0.1-1.0$
        respectively. In order to obtain better scaling exponent, only for the case $\St=0.17$, $\phim=1$, as
        $\dtwo$ is large,  we take $\Np=4.5\cdot10^{5}$.} 
	\end{center}
\end{figure}

\section{\label{model} Model and Numerical Method}

The dust is modeled as a system of mono-dispersed spherical particles governed by \Eq{eq:HIP}.
Gas is modeled in the Eulerian-framework where the equation for the scalar vorticity field $\omega(\bm{x},t) \equiv \nabla\times\bm{u}(\bm{x},t)$ is
\begin{equation}
  D_t \omega(\bm{x},t) = \nu\nabla^2\omega(\bm{x},t) - \alpha \omega(\bm{x},t) + f(\bm{x},t) + 
    \nabla\times \Fdtog(\xx,t)\/.
  \label{eq_ns_vel}
\end{equation}
Here $\uu(\xx, t)$ is the incompressible velocity field, 
$D_t = \partial_t + \bm{u\cdot}\nabla$ 
is the material derivative, $\nu$ is the viscosity, $\alpha$ is the Ekman drag coefficient, 
and 
$f(\xx,t) =   -\fnot \kf \cos(\kf  y)$ 
is the Kolmogorov forcing with amplitude $\fnot$ and at wave-number $\kf$. 
The force exerted by the dust particles on the  gas is
\begin{equation}
\Fdtog(\xx,t) = \sum_{{\rm i}=1}^{N_{\rm p}} \frac{m}{\taup \rhog}\left[\VV_{\rm i} - \uu(\xx,t)\right]
         \delta^2(\xx-\XX_{\rm i}),
\label{eq:force}
\end{equation}
where $m$ is the mass of a dust particle and $\Np$ is the total number of particles.
We use a pseudo-spectral method \cite{Can88,per+ray+mit+pan11} to numerically 
integrate \Eq{eq_ns_vel} in a periodic square box with each side of length $L = 2\pi$. 
The simulation domain is spatially discretized using $N^2$ collocation points.  
For time evolution we employ a second-order Runge-Kutta scheme \cite{Pre+Fla+Teu+Vet92}.

In this Eulerian-Lagrangian framework, the position of a dust grain does not, in general, 
coincides with the Eulerian grids. 
The gas velocity at the position of a dust in \Eq{eq:dvdt} is obtained as 
\begin{equation}
\uu(\XX,t) = \sum_{\xx} {\bm u}(\xx,t) \delta^2_{\rm h}(\xx - \XX) h^2\/,
\end{equation}
where $h=L/N$, and the $\delta^2_h(\cdot)$ is 
a numerical realization of the two-dimensional delta function
on grids of linear dimension $h$.  
We use the following prescription~\cite{peskin}:
\begin{equation}
 \delta(x-X) =\begin{dcases}
 \frac{1}{4h}\bigg\{1+\cos\left[\frac{\pi (x - X)}{2h}\right]\bigg\}, \ \ \ \ |x-X| \leq 2h,\\
 \   0 \ \ \ \ \ \text{otherwise}.
 \end{dcases}
 \label{cos_scheme}
\end{equation}
The same prescription, \Eq{cos_scheme}, is also used to discretize the delta function 
in \Eq{eq:force}. We initialize our simulation with $\Np$ randomly placed dust particles. 
Similar to aerosols in clouds \cite{sha03}, we assume $\rhod/\rhog \sim 10^3$. 
The vorticity is initialized as 
$\omega(\bm{x},0) = -\fnot \kf \nu\left[ \cos(\kf x) + \cos(\kf y)\right]$. 
\section{\label{result} Results}
We study the dust-gas turbulence 
by varying the mass-loading parameter 
$\phim \equiv {\Np m/(\rhog L^2)}$ and the Stokes number $\St$.

\subsection{\label{clustering} Vorticity and Clustering}

\begin{figure}[!h]
	\includegraphics[width=0.4\textwidth]{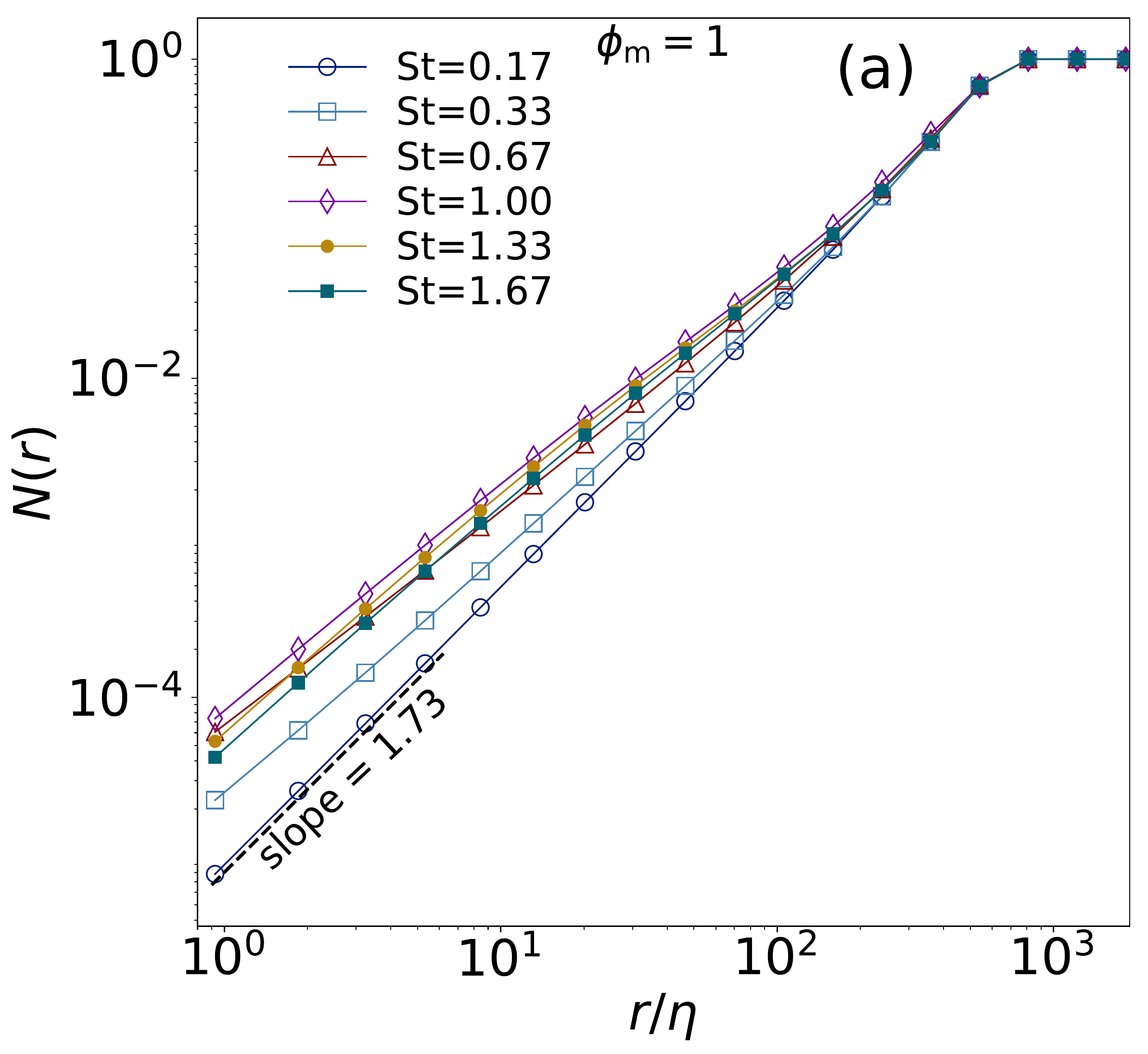}
	\includegraphics[width=0.4\textwidth]{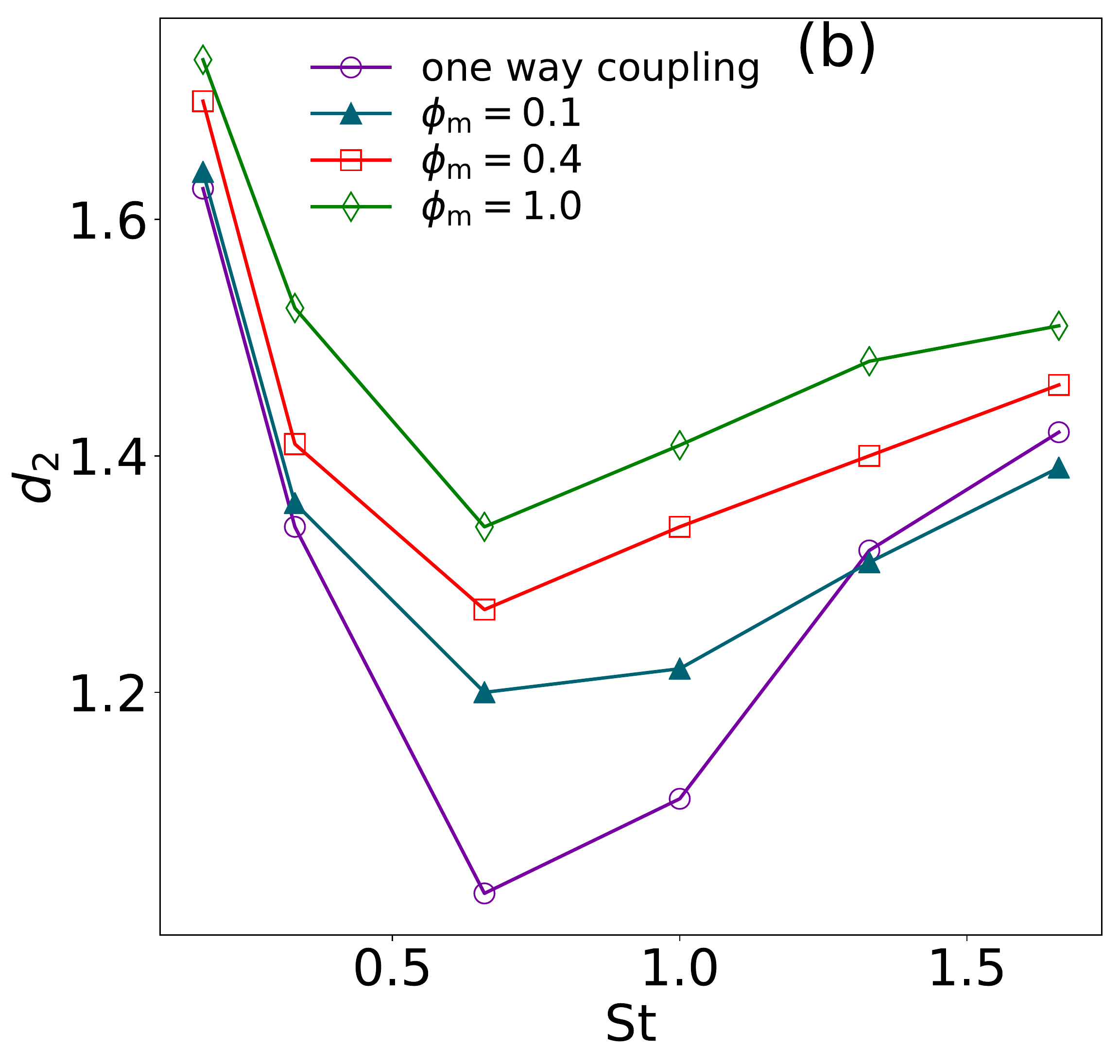}
	\caption{\label{fig_corr} (a) Log-log plot of the cumulative distribution function 
		$N(r)$ vs. $r/\eta$ for $\phim=1$ and different $\St$. The dashed line represents 
		least square fit with corresponding slope value.
		(b) The correlation dimension $\dtwo$ vs.  $\St$ for different $\phim$.
		}
\end{figure}

In \fig{fig:vort} we show the pseudo-color plot of the vorticity field in absence of 
dust ($\phim=0$) and  at high mass-loading $\phim=1$. 
We observe that the in the latter small-scale vortices form in the regions where particle 
cluster. 
Our observation is consistent with the earlier study of two-dimensional dusty-gas 
turbulence~\cite{bec_dustygas}.  
We quantify the clustering by using the cumulative pair distribution function
\begin{equation}
	{N}(r) \equiv \left\langle {2\over \Np(\Np-1)}\sum_{i<j} \Theta(r-\mid{\mathbf X}_{\rm i}-
             {\mathbf X}_{\rm j}\mid)\right\rangle.
\end{equation}
Here $\Theta$ is the Heaviside function and the angular brackets denote averaging over different 
stationary-state turbulent configurations. 
In \subfig{fig_corr}{a}, we plot $N(r)$ versus $r$  
for fixed $\phim=1$ and with different $\St$. In the limit $r{\to 0}$, $N(r) \sim r^{\dtwo}$, 
where $\dtwo$ is the correlation dimension \cite{procaccia_d2}. 
We obtain $\dtwo$ by performing a least square fit in the range $1<r\eta<10$. 

In \subfig{fig_corr}{b} we plot the correlation dimension $\dtwo$ as a function of  $\St$ 
for different values of $\phim$.  Note that $\dtwo=2$ for $\St=0, \infty$ and 
attains a minimum value, which corresponds to maximum clustering, around 
$\St\approx 0.6$ \cite{bec+bif_2007}. 
We observe that for all the values of $\phim$ this is indeed the case. 
However, the amount of clustering (smallest value of $\dtwo$) decreases with 
increasing $\phim$. 
We find that for  a fixed $\St$, the maximum clustering is obtained for one-way coupled 
simulations where the back-reaction from  the dust  is ignored.  
Similar results have also been observed for particle-laden turbulent homogeneous shear 
flows~\cite{gua_2011,paolo_2017}.
Qualitatively,  the small-scale vortices produced in presence of mass-loading 
will expel particles hence clustering reduces as  with $\phim$ increases.

\subsection{\label{spectra} Energy spectra}
\begin{figure}[!h]
	\includegraphics[width=0.8\linewidth]{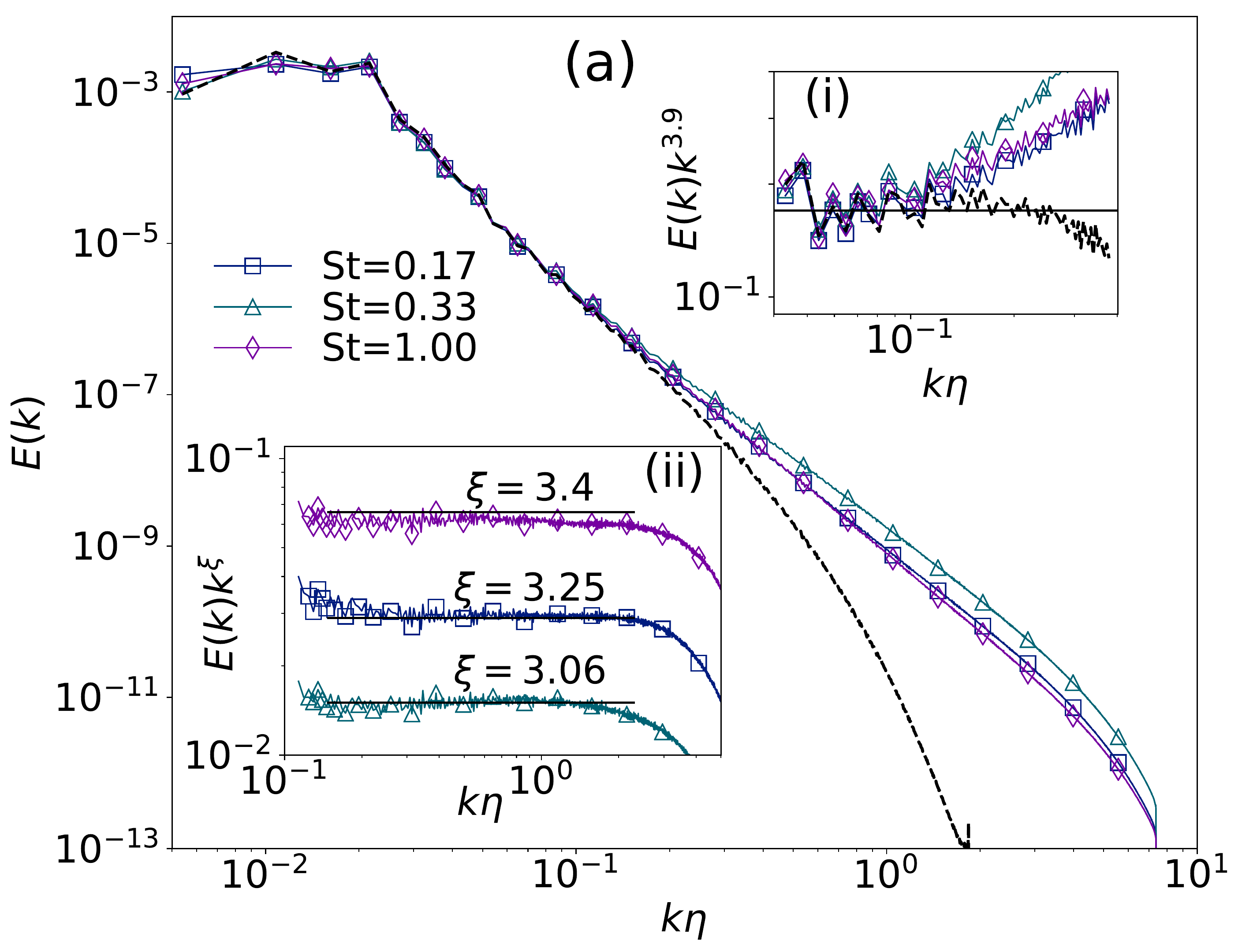} 
	\includegraphics[width=0.8\linewidth]{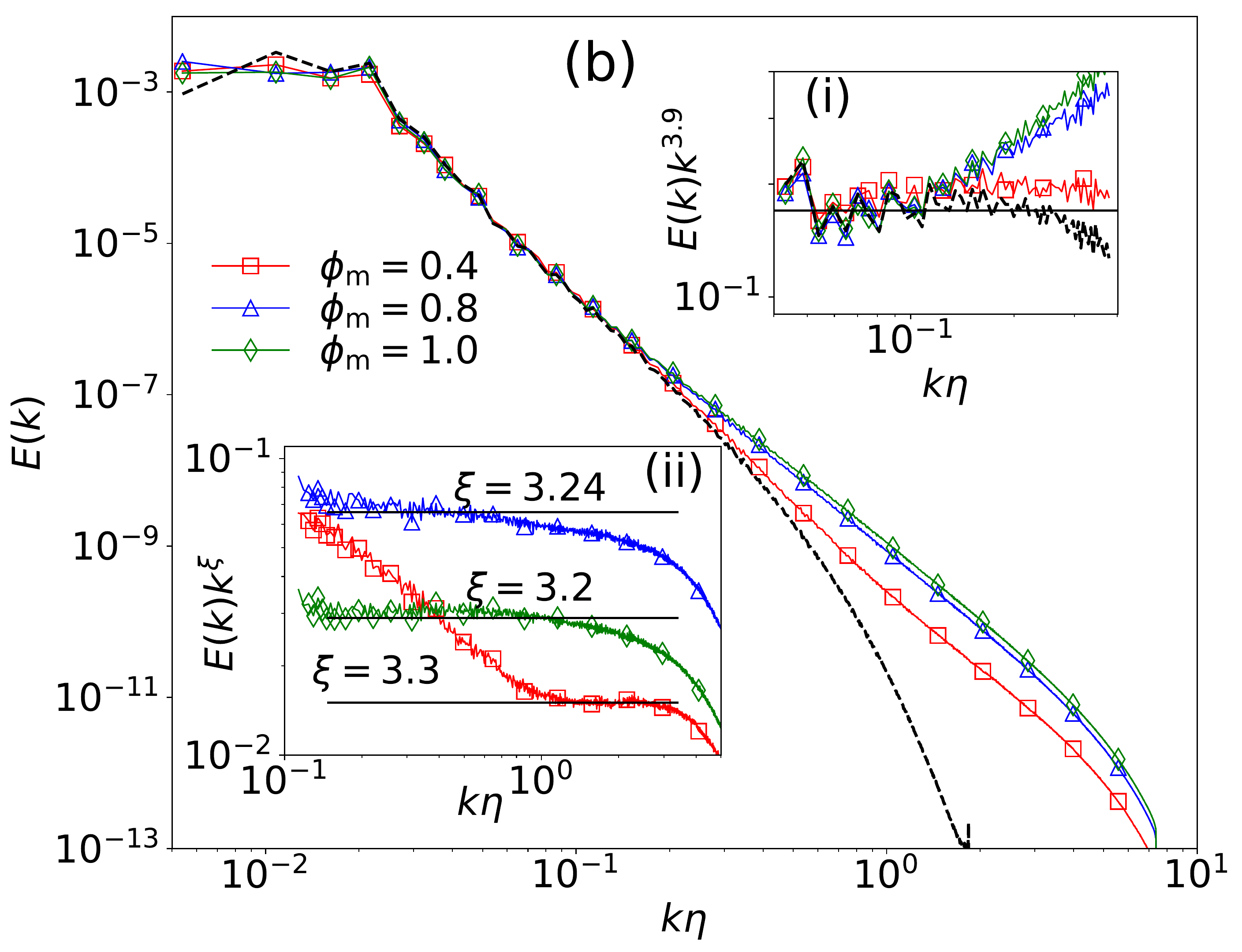} 
	\caption{\label{fig:spectra} Log-log plot of spectra
		 for (a) $\phim = 1.0$ fixed, $\St$ varied, (b) $\St=0.67$ fixed and $\phim$
          varied.
           Black dashed line represents $E(k)$ for  $\phim=0$.  
           [Inset (i)] Log-log plot of energy spectra
          $E(k)$ compensated by $k^{3.9}$. Note that the dusty-gas spectrum 
          deviates from the $\phim=0$ case, which is marked by the rise in tail.
          [Inset (ii)] Log-log plot of energy spectra compensated by
        $k^{\xi}$ where the exponents $\xi$ for each $\St$ and $\phim$ is 
        given in the inset.} 
\end{figure}

 Next, we study the angle averaged velocity power spectrum  
\begin{equation}
E(k) \equiv {1\over 2}\left\langle\sum_{k-1/2 \leq m < k+1/2} 
\mid\bm{u}_{\bm{m}}\mid ^2\right\rangle\/,
\end{equation}
 where $\bm{u}_{\bm{m}}$ is the velocity field in the Fourier space. 
In the absence of dust particles ($\phim=0$), steady-state two-dimensional energy 
spectrum [\subfig{fig:vort}{a}]  shows inertial range scaling 
$E(k)\sim k^{-3.9}$ for $0.03 \leq k \eta \leq 0.1$ 
and decays exponentially in the dissipation range ($k\eta >0.10$)~\cite{per+ray+mit+pan11}. 
We find that addition of dust particles to the gas dramatically alters  the dissipation 
range spectrum. For $k\eta>0.1$, we observe a new power-law 
$E(k)\sim k^{-\xi}$ with  $\xi<3.9$. 
Let us hold $\phim=1$ fixed and increase $\St$ (\subfig{fig:spectra}{a}):
 we find $\xi$ first decreases, reaches its minimum value $\xi\sim3$ for $\St=0.33$ 
[see insets to \subfig{fig:spectra}{a}] and then increases again. 
For a fixed $\St=0.67$, $\xi$ reduces monotonically as $\phim$ is increased \subfig{fig:spectra}{b}. 
   
\subsection{\label{budget} Enstrophy budget}
To understand the scaling behavior we now study the scale-by-scale 
enstrophy budget equation:
\begin{equation}
\Pi(k) = \mathscr{D}(k) - \alpha \Omega(k) + \mathscr{F}(k) + \mathscr{R}(k).
\label{ns_ener_bud}
\end{equation}
Here 
\begin{equation}
\nonumber
\Omega(k) \equiv \left\langle\sum_{m \leq k}\mid\omega_m\mid^2\right\rangle
\end{equation}
 is the cumulative enstrophy up to wave-number $k$, 
\begin{equation}
\nonumber 
\Pi(k) \equiv \left \langle\sum_{m \leq k}
\omega_{\bm{m}}(\bm{u\cdot}\nabla\bm{\omega})_{-\bm m} \right \rangle
\end{equation}
 is the enstrophy flux due to non-linear terms, 
\begin{equation}
\nonumber 
\mathscr{D}(k) \equiv -\nu \left \langle \sum_{m \leq k} m^2\mid\omega_{\bm{m}}\mid^2 \right\rangle\/,
\end{equation}
 is cumulative dissipation rate, $-\alpha\Omega(k)$ is 
the contribution due Ekman friction, 
\begin{equation}
\nonumber 
\mathscr{F}(k) \equiv \left\langle \sum_{m \leq k} \omega_{\bm{m}} 
      f_{-\bm{m}}\right\rangle
\end{equation} 
is the cumulative energy injected due to Kolmogorov forcing, and 
\begin{equation}
\nonumber 
\mathscr{R}(k) \equiv \left \langle \sum_{m \leq k} \omega_{\bm{m}}(\nabla\times 
           \Fdtog)_{-\bm m}\right \rangle
\end{equation}
 is the contribution because of back reaction from the dust particles to the gas.
\begin{figure}
\includegraphics[width=0.8\linewidth]{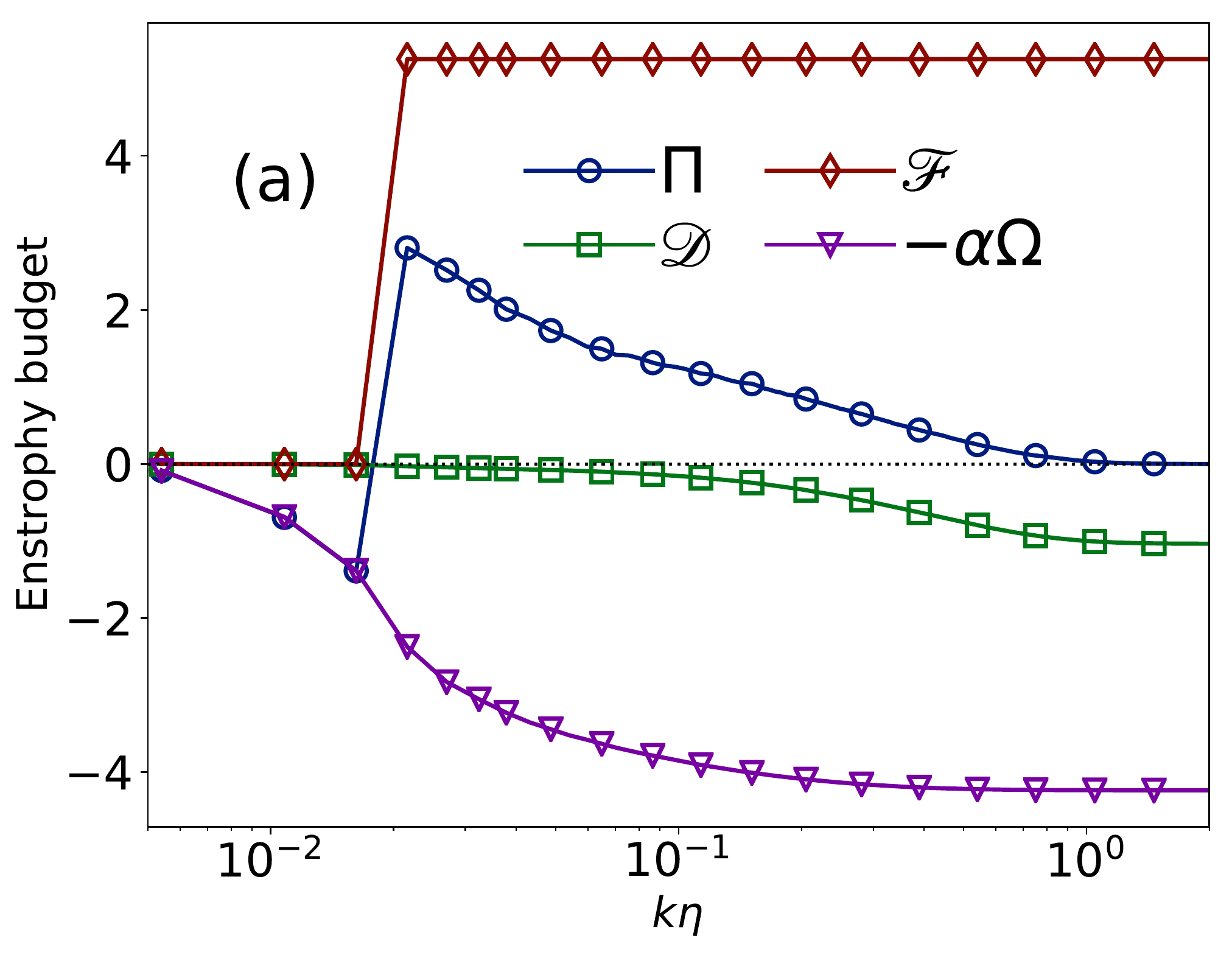}
\includegraphics[width=0.8\linewidth]{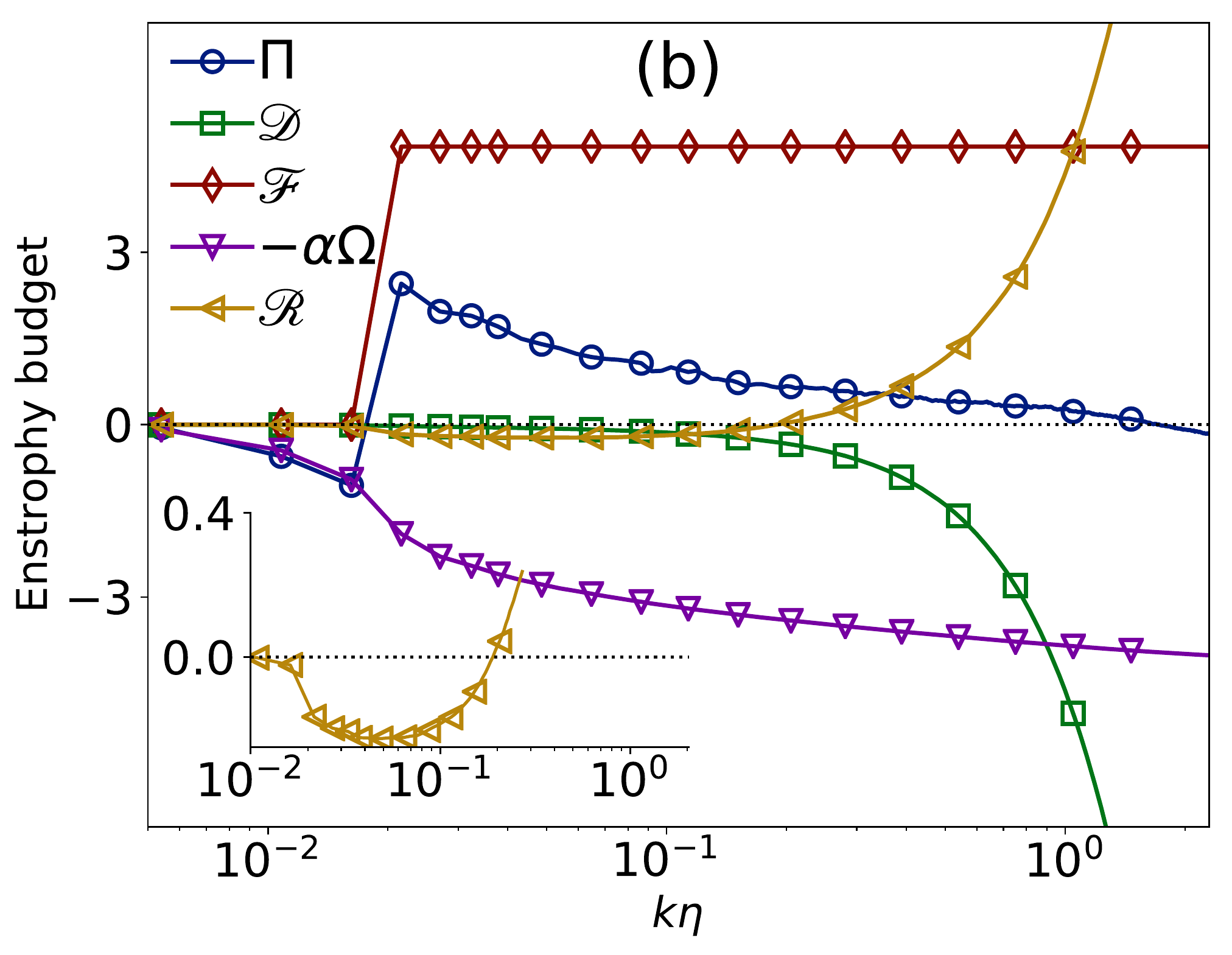}
\caption{\label{bal_0} Semi-log (X-axis in log scale) plot of scale-by-scale enstrophy budget
    for (a) $\phim=0$, (b) $\St=0.67$ and $\phim = 1$. (Inset) Semi-log plot of $\mathscr{R}(k)$ zoomed 
     for $k\eta
    \leq 0.3$. Note that $\mathscr{R}(k)$ is negative till wave mode $k_c$.
    Y-axis in both (a) and (b) is normalized by enstrophy dissipation rate $\beta$.}
\end{figure}
In \subfig{bal_0}{a} we plot the enstrophy budget for the gas in absence of 
particles ($\phim = 0$).  Similar to earlier studies, we observe that at large 
scales energy injected by external forcing is primarily balanced by Ekman drag and 
the enstrophy flux $\Pi(k)$ decreases with increasing $k$ \cite{bof+eck12}.

\begin{figure}[!ht]
	\includegraphics[width=0.8\linewidth]{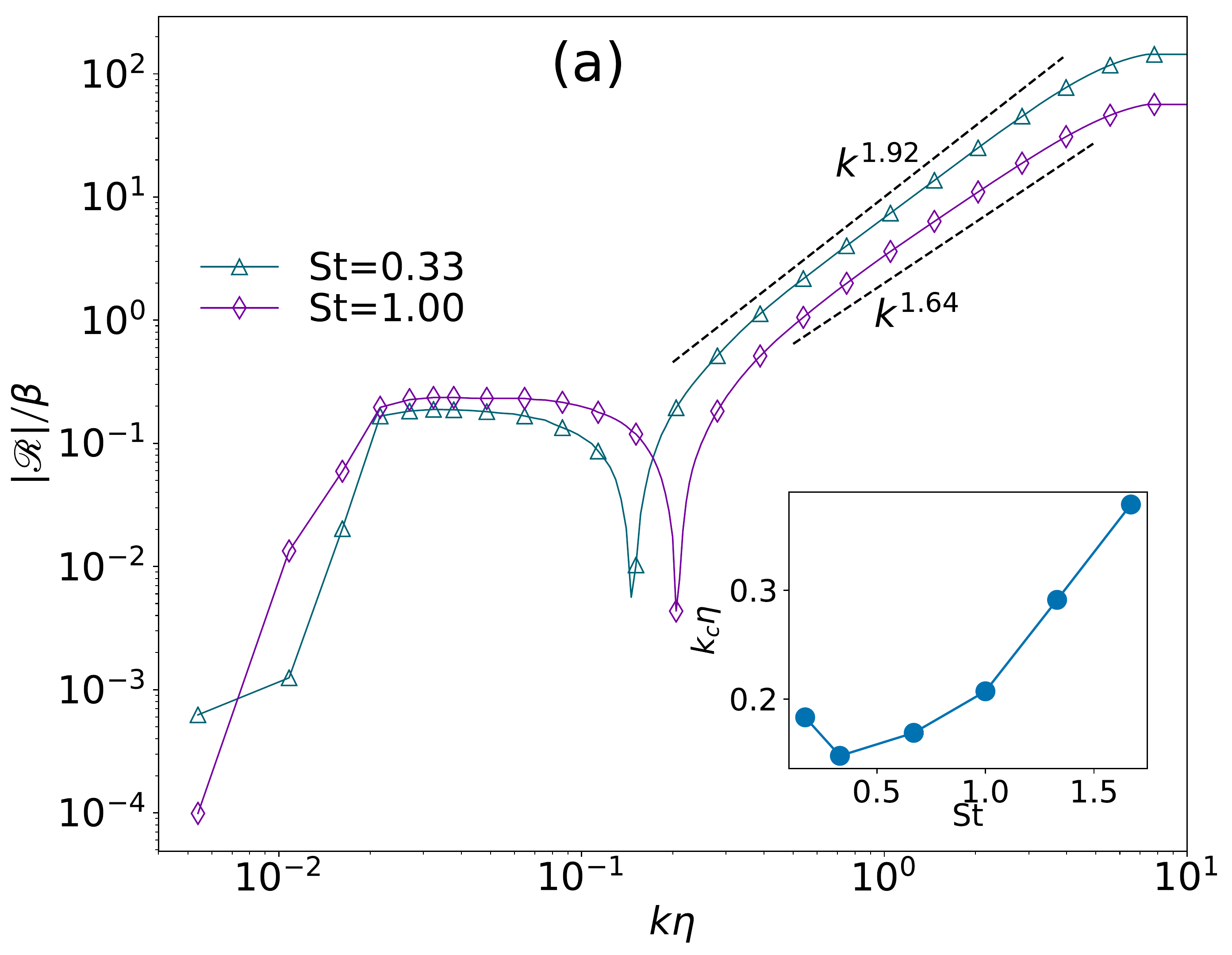}
	\includegraphics[width=0.8\linewidth]{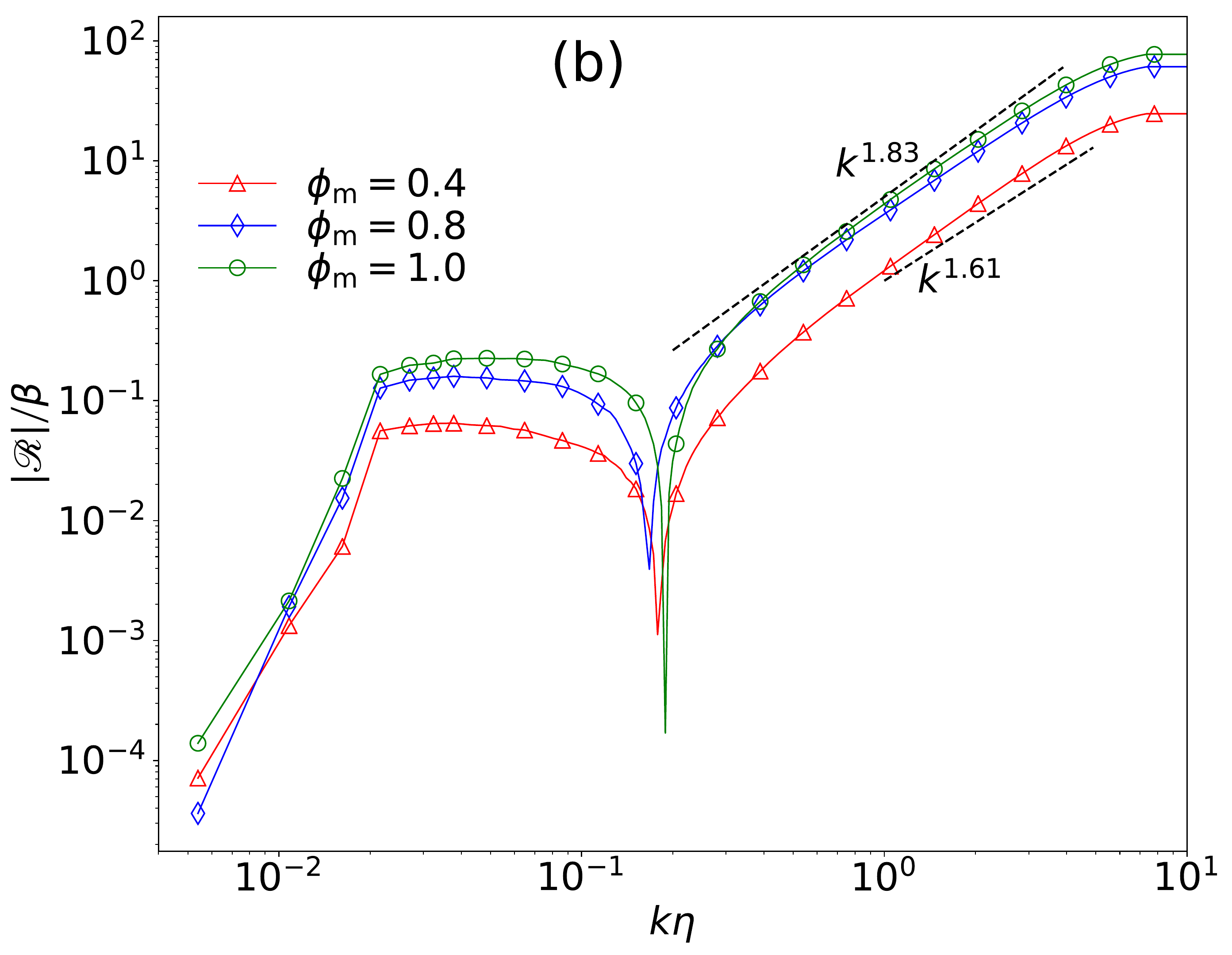}
	\caption{\label{ens_bud_2} Log-log plot of $\mathscr{R}(k)$ for (a) $\phim = 1$ fixed, $\St$
    varied and (b) $\St=0.67$ fixed and $\phim$ varied. (Inset) Linear plot of $\kc\eta$  (mode
    where $\mathscr{R}(k)$ changes sign) vs.
    $\St$. Dashed lines in both (a) and (b) shows the 
    scaling for $\mathscr{R}(k)$.}
\end{figure}

We now show that the presence of dust particles dramatically alters the enstrophy budget in 
the dissipation range.  In \subfig{bal_0}{b} we plot the cumulative contributions
of all the terms in budget for  $\St=0.67$ and $\phim = 1.0$. 
The dust particles inject enstrophy ($\mathscr{R}$) at  large $k$ which is then  balanced by 
viscous dissipation $\mathscr{D}$. We find a negligible change in shape of $\Pi$, 
$\mathscr{F}$ and the Ekman drag term in the inertial range. 
A closer look at $\mathscr{R}$ [inset \subfig{bal_0}{b}] reveals that it makes a net negative 
contribution to budget till a  wavenumber $\kc$ after which it turns positive.  
Clearly, the particles extracts enstrophy from the flow at small $k$ (large scales)  but 
injects enstrophy at large $k$ (small scales).  
Furthermore, for $k>\kc$ the two dominant terms that balance each other
are  $\mathscr{R}$ and $\mathscr{D}$.  Hence, we expect  
$E(k) \sim k^{-4} d[\mathscr{R}(k)]/dk$ for $k>\kc$. 
In \subfig{ens_bud_2}{a,b} we show that  
$\mathscr{R}(k)\sim k^{\beta}$ for $k>\kc$.  
Using the relation between $E(k)$ and $R(k)$, we find the scaling exponent 
$\xi=-5+\beta$. For $\phim=1$ and $\St=0.33, 0.67, 1.0$ we obtain 
$\xi =3.08, 3.17 ~\rm{and} ~3.36 $ respectively, 
which is very close to the exponents that we directly determine from the energy 
spectra [\subfig{fig:spectra}{}]. This proves that the dominant balance between
$\mathscr{R}$ and $\mathscr{D}$ determines the new scaling exponent.  

\section{\label{conclusion}  Conclusion}
We use an Eulerian-Lagrangian formalism to study the effects of dust to gas
coupling in two-dimensional turbulence.
The dust are modeled as heavy inertial particles immersed
in the gas. We solve gas equations on fixed Eulerian grids by incorporating 
the forces [\Eq{eq:force}] due to dust . The main problem with this 
technique is that to have a smooth Eulerian
representation of the feedback, number of particles per cell
needs to be equal or greater than a certain 
threshold ($\approx 1$) \cite{balchandar_rev_2010,paolo_2013}. 
We choose $\Np$ such
that in the stationary state, i.e. after the dust have clustered, 
the ratio $\Np/N^{\dtwo} \approx 1$ for almost all the $\St$.
Furthermore, we use higher order weight function 
for extrapolation to ensure a better approximation of back reaction 
in fluid grids.
We obtain reasonable scaling range for nearly all 
the $\St$ and $\phim$. 

The Eulerian-Lagrangian formalism has been extensively used
to understand how the interaction with dust modifies
three-dimensional turbulence. 
Here, we shall review some of them with 
an emphasis on energy spectra (see
\cite{balchandar_rev_2010,poema06} and references therein for more details). 
Refs.~\cite{eaton_1990,boivin_1998} studied the effects of dust in 
isotropic stationary turbulence using direct numerical simulations while
similar studies in decaying turbulence were done by Refs. \cite{ferrante_2003,sun_col_1999}.
The key results of these studies are: (a) particle injects 
energy at large $k$ and reduces it at small $k$, and (b) increasing mass loading 
leads to reduction of the total kinetic energy.  But, the effect of $\St$ or $\phim$ on the scaling of energy 
spectra remained unclear as these simulations were done at small or moderate resolution.

More recently, Ref. \cite{gua+pic+sar+cas15} introduced a new 
numerical scheme to model coupling between gas and dust wherein the disturbance in the gas due to 
dust particle is evaluated in a closed analytic form~(by using solutions of the Stokes equation) 
and is incorporated into fluid grids after a certain regularization time ($\epsilon_{\rm r}$). Here $\epsilon_r$ is the
time taken by a subgrid-scale disturbance to reach nearby fluid grid locations from an off-grid particle location. 
One major advantage of this method is that the number of particles need not be comparable to number of grid cells for smooth feedback. Unfortunately, the method is computationally expensive and not 
easily parallelizable on distributed-memory machines.  By studying dust laden homogeneous shear turbulent flow using this technique, Ref. \cite{gua+bat+cas17} reported a scaling exponent of $-4$ in gas kinetic energy spectrum [for $\St=1$ and $\phim=(0.4,0.8)$]. They argued that the new scaling appears due to the balance of viscous forces with the back reaction from dust.  In \cite{gua+bat+cas17}, the new scaling was observed for $k\eta \geq 1$, whereas our two-dimensional study shows that it starts around  $k\eta \sim 0.2$. Clearly, the crucial problem with our and similar studies is that there is, as yet, no well-established algorithm to numerically calculate the feedback in DNS.  

For good reason, the most important one being difficulties in experimental realization, 
turbulence in flows of dust and gas has been rarely studied in two dimensions. 
Ref.~\cite{bec_dustygas} using Eulerian description of dust found a 
scaling exponent of $-2$ in the gas energy spectra, that once again
emerges due to  balance of viscous dissipation  against the feedback, 
for $\St \ll 1$ and $\phim$ between $0.1-0.4$.
To numerically smoothen the caustics that invariably develops 
in such a computation a synthetic hyper-viscous term was added in the
Eulerian description. 
Ekman drag coefficient was chosen such that the pure gas spectra
(without dust coupling) scale with an exponent $-3.3$. Notably, the 
new scaling here starts at much small $k$ compared to what we find. 

To summarize our main results: (a) presence of dust-gas coupling decreases clustering
of dust particles, (b) a new scaling regime emerges in the kinetic energy spectrum, 
marked by rise in the tail, (c) scale-by-scale enstrophy budget, suggests that the new scaling 
is because of gas viscosity dissipating the enstrophy injected by dust at those scales. 
(d) dust has a net negative contribution to budget till a wavenumber 
 $\kc$ and injects enstrophy at higher fourier modes,
(e) as the form of dust-gas coupling term varies with 
both $\phim$ and more importantly $\St$, the scaling exponent is 
non-universal and a function of both. 

We conclude that once feedback from the dust to the gas is significant
the spectra of gas changes, not in the inertial range but a new
scaling regime emerges in the erstwhile dissipative range. 
The central message of this paper is that this scaling exponent is non-universal,
it depends on the Stokes number of the dust and the mass loading parameter. 
Even in two dimensions, where we have been able to do large-scale simulations 
for long enough time, the appearance of the new scaling regime is not always 
prominent. We cannot rule out the possibility that there may be no scaling range 
at all. But we can and do conclude that the spectra is \textit{non-universal}. 

It is quite difficult to perform a DNS of similar resolution, with feedback from particles,
in three dimensions. So it is unlikely that in near future we shall observer clear scaling 
behavior in analogous cases in three dimensions. But based on our result we speculate that
the same \textit{non-universal} nature of spectra will be true in three-dimensions too. 

\section{Acknowledgment}
We thank Paolo Gualtieri for useful discussions. 
DM acknowledges financial support from the grant Bottlenecks
for particle growth in turbulent aerosols from
the Knut and Alice Wallenberg Foundation (Dnr. KAW
2014.0048) and from Swedish Research Council Grant no. 638-2013-9243 as well as 
2016-05225. 
\end{document}